\documentclass{elsart5p}

\usepackage{amsfonts,amssymb,amsmath}
\usepackage{bm}
\usepackage[ansinew]{inputenc}
\usepackage[english]{babel}
\usepackage[OT1]{fontenc}
\usepackage{graphicx}

\newcommand{\Na}[1]{Na$_{#1}$CoO$_2$}
\newcommand{\vect}[1]{\bm{#1}}

\begin{document}

\begin{frontmatter}

\title{Quasiparticle dispersion near the Fermi surface in \Na{x}}

\author[LPS]{A.~Bourgeois},
\author[CAB]{A.A.~Aligia},
\author[IFW]{T.~Kroll},
\author[LPS]{M.D.~N\'u\~nez-Regueiro}
\address[LPS]{Laboratoire de Physique des Solides, CNRS UMR-8502, Universit\'e Paris-Sud, 91405 Orsay cedex, France}
\address[CAB]{Comisi\'on Nacional de Energ\'ia At\'omica, Centro At\'omico Bariloche and Instituto Balseiro, 8400 S.~C. de Bariloche, Argentina}
\address[IFW]{IFW Dresden, P.O. Box 270016, D-01171 Dresden, Germany}

\begin{abstract}

We construct an effective Hamiltonian for the motion of $t_{2g}$ highly correlated states in \Na{x}. This three-band model includes the indirect 
Co--O--Co hopping $t$ and the crystal-field splitting $3D$. Calculations in a CoO$_6$ cluster give the effective parameters $t = 100$~meV and $3D = 315$~meV. The Hamiltonian is solved using a generalized slave-boson mean-field approximation. The results show a significant band renormalization. Without any additional hypothesis, the dispersion of the bands near the Fermi energy and Fermi surface topology agrees with angle-resolved photoemission experiments, in contrast to predictions using the local-density approximation.

\end{abstract}


\end{frontmatter}

\section{Introduction}
\label{Sec:Introduction}

The doped layered hexagonal cobaltates \Na{x} have attracted considerable interest in the last few years since the discovery of a superconducting phase (\Na{0.35}--1.3~H$_{2}$O)~\cite{NAT42253} and of an exceptionally high thermopower~\cite{NAT423425}. The crystallographic structure of \Na{x} consists in electronically active layers of edge-shared CoO$_6$ octahedra alternating with layers of Na. The physics of these compounds is believed to be essentially two dimensional because of the weak coupling between CoO$_2$ layers, and Na atoms act as a charge reservoir whose order can dramatically affect the CoO$_2$ planes like in \Na{0.5}~\cite{PRL96107201}. The knowledge of the band structure and the Fermi surface (FS) of these systems is crucial to understand their physical properties. However, while local-density approximation (LDA) calculations predict a FS which consists of a large portion around the $\varGamma$ point of the Brillouin zone, and six pockets near the $K$ points~\cite{PRB70045104}, angle-resolved photoelectron spectroscopy (ARPES) revealed only a single hole like FS around $\varGamma$ and no hole pockets~\cite{PRL95146401}. Whether a single-band model can properly describe these compounds is still under debate. Recent experiments with X-ray spectroscopy (XAS) and their interpretation~\cite{PRL94146402,Kroll1,Kroll2} show a strong Co--O covalency which should be taken into account. Here we study a multiband Hamiltonian with just two ingredients which we believe are the essential ones: the indirect hoppings between the three $t_{2g}$ orbitals through O $2p$ orbitals and the trigonal splitting. Both are derived from cluster calculations which take properly Co--O covalency and correlations into account.

\section{The multiband model}
\label{Sec:Model}

\begin{figure}[htbp]
	\centering \includegraphics[width=.7\linewidth]{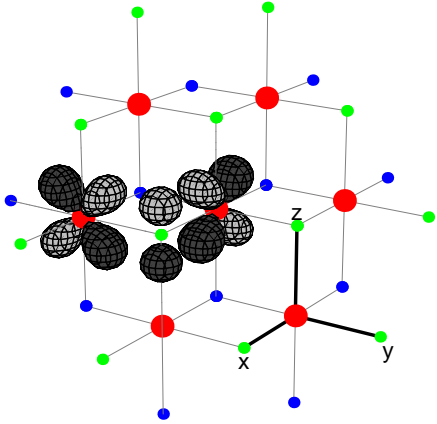}
	\caption{(Color online) Orbitals involved in the effective hopping of a hole with symmetry $yz$ to a neighboring state with symmetry $zx$, through a $p_{z}$ orbital of an intermediate O site.}
	\label{Fig:Hopping}
\end{figure}

As first pointed out by Koshibae and Maekawa~\cite{PRL91257003}, the dominant hopping process for a carrier on a given Co $t_{2g}$ orbital to specific nearest Co orbitals takes place through an intermediate O $2p$ orbital. See Fig.~\ref{Fig:Hopping}. Starting from a given $t_{2g}$ orbital, the states accessible by successive indirect hoppings form a Kagom\'e sublattice. In the basis $\bigl\{ d_{xy}, d_{yz}, d_{zx} \bigr\}$ of the $t_{2g}$ orbitals, the indirect hopping matrix reads~\cite{PRL91257003}
\begin{equation*}
	\epsilon_{\gamma\gamma'}(\vect{k}) = 2t
		\begin{pmatrix}
		0 & \cos \theta_{3} & \cos \theta_{2} \\
		\cos \theta_{3} & 0 & \cos \theta_{1} \\
		\cos \theta_{2} & \cos \theta_{1} & 0
		\end{pmatrix},
\end{equation*}
with $\theta_{1} = k_{x} a$, $\theta_{2} = \frac{k_{x} a}{2} - \frac{\sqrt{3} k_{y} a}{2}$ and $\theta_{3} = \frac{k_{x} a}{2} + \frac{\sqrt{3} k_{y} a}{2}$, where here $k_x$ refers to the direction of two nearest neighbors Co atoms ((-1,1,0) in Fig.~\ref{Fig:Hopping}). The trigonal distortion in the (1,1,1) direction splits the $t_{2g}$ orbitals into a doublet $e'_g$ and a singlet $a_{1g}$ with $E(a_{1g}) - E(e'_g) = 3D$. This corresponds to an on-site potential
\begin{equation*}
	D_{\gamma \gamma'} = D
		\begin{pmatrix}
		0 & 1 & 1 \\
		1 & 0 & 1 \\
		1 & 1 & 0
		\end{pmatrix}.
\end{equation*}
Calculations on a CoO$_6$ cluster~\cite{Kroll2,CM0605454}, in which correlations were treated properly, lead to a strongly covalent system and give 
$t = 100$~meV, $3D = 315$~meV. Thus we have $D \sim t$, therefore the splitting is not strong enough to justify a single-band model based on localized $a_{1g}$ orbitals. 

The resulting effective model for the motion of the doped holes in the system can be written in the form
\begin{equation*}
	\begin{split}
		H_{eff} &= \sum_{\vect{k}} \sum_{\gamma,\gamma',\sigma} \bigl( \epsilon_{\gamma\gamma'}(\vect{k}) + D_{\gamma\gamma'} \bigr) h^{\dag}_{\vect{k}\gamma\sigma} h^{}_{\vect{k}\gamma'\sigma} \\
			&\quad + U \sum_{\vect{k}} \sum_{\gamma} n^{}_{\vect{k}\gamma\uparrow} n^{}_{\vect{k}\gamma\downarrow}
	\end{split}
	\label{Eq:Heff}
\end{equation*}
with $U \rightarrow \infty$ to avoid double occupancy. This strong-interaction limit is justified by comparison of the energy scale $D \sim t$ with the remaining correlations in the effective model. The effective model is solved using the slave boson treatment of Kotliar and Ruckenstein in mean-field~\cite{PRL571362} (which is equivalent to the Gutzwiller approximation) generalized to the multiband case~\cite{PRB475095}, resulting in a renormalization of the hoppings.

\section{Results}
\label{Sec:Results}

\begin{figure}[htbp]
	\centering
	\includegraphics[width=.6\linewidth]{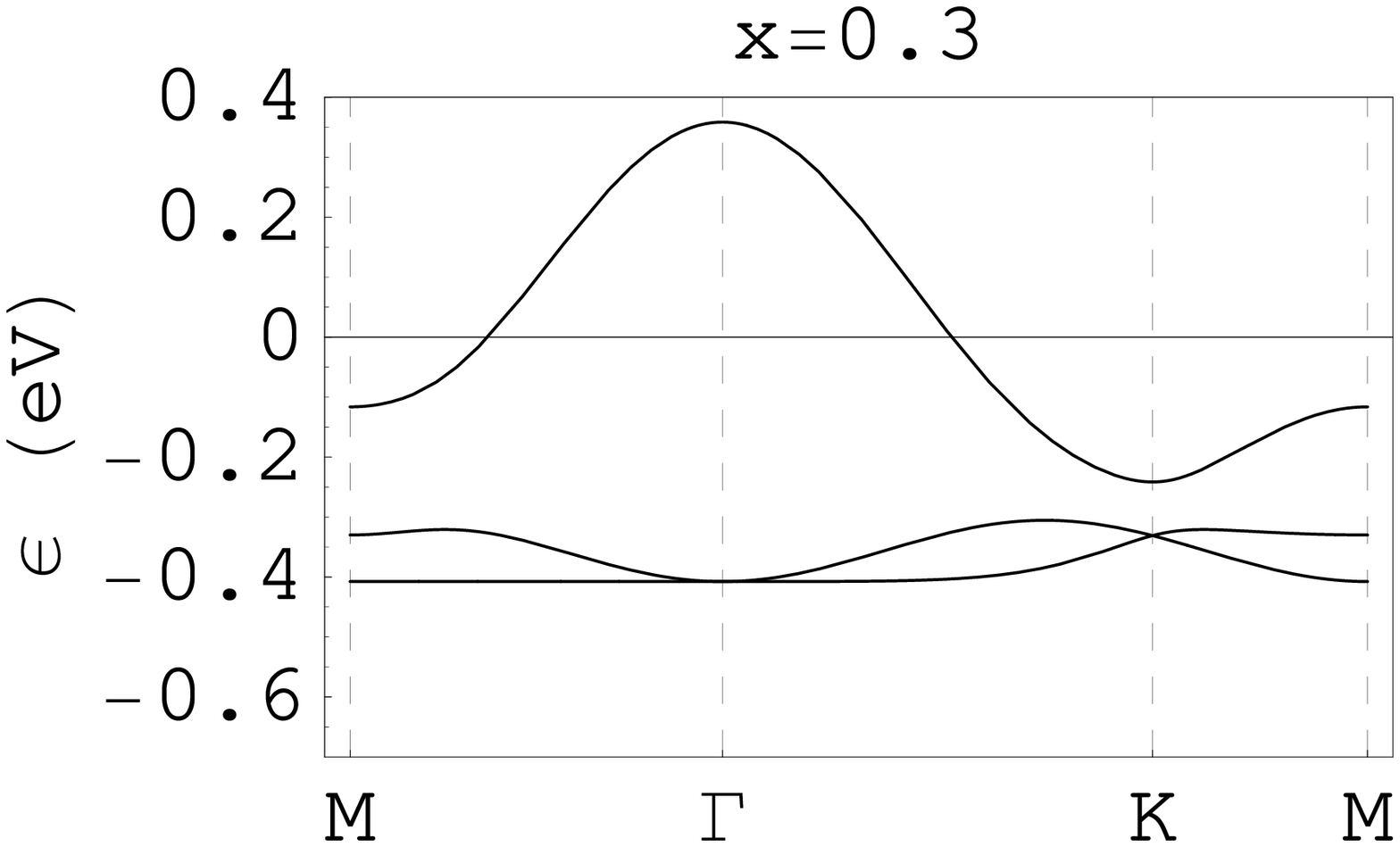}
	\includegraphics[width=.6\linewidth]{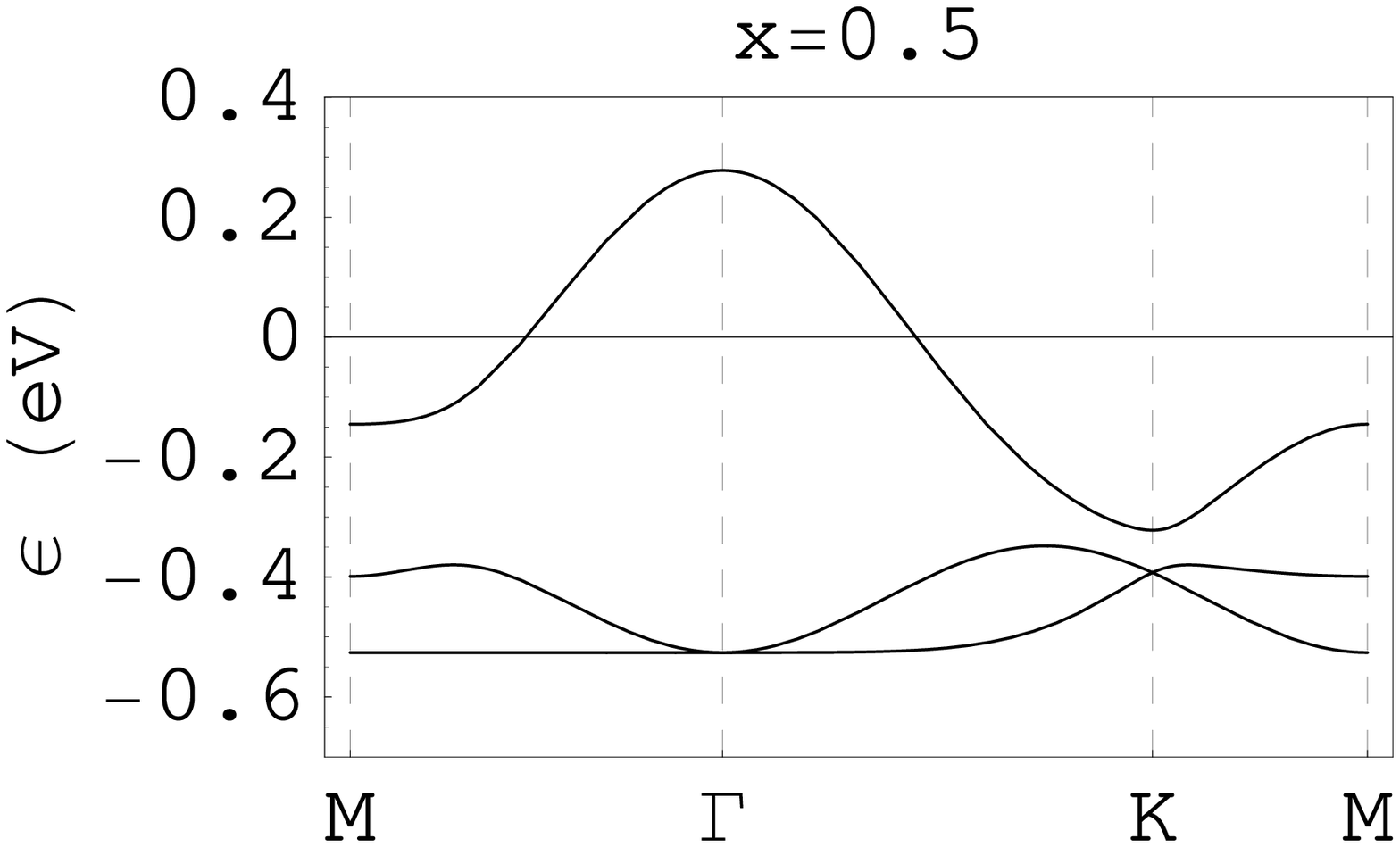}
	\includegraphics[width=.6\linewidth]{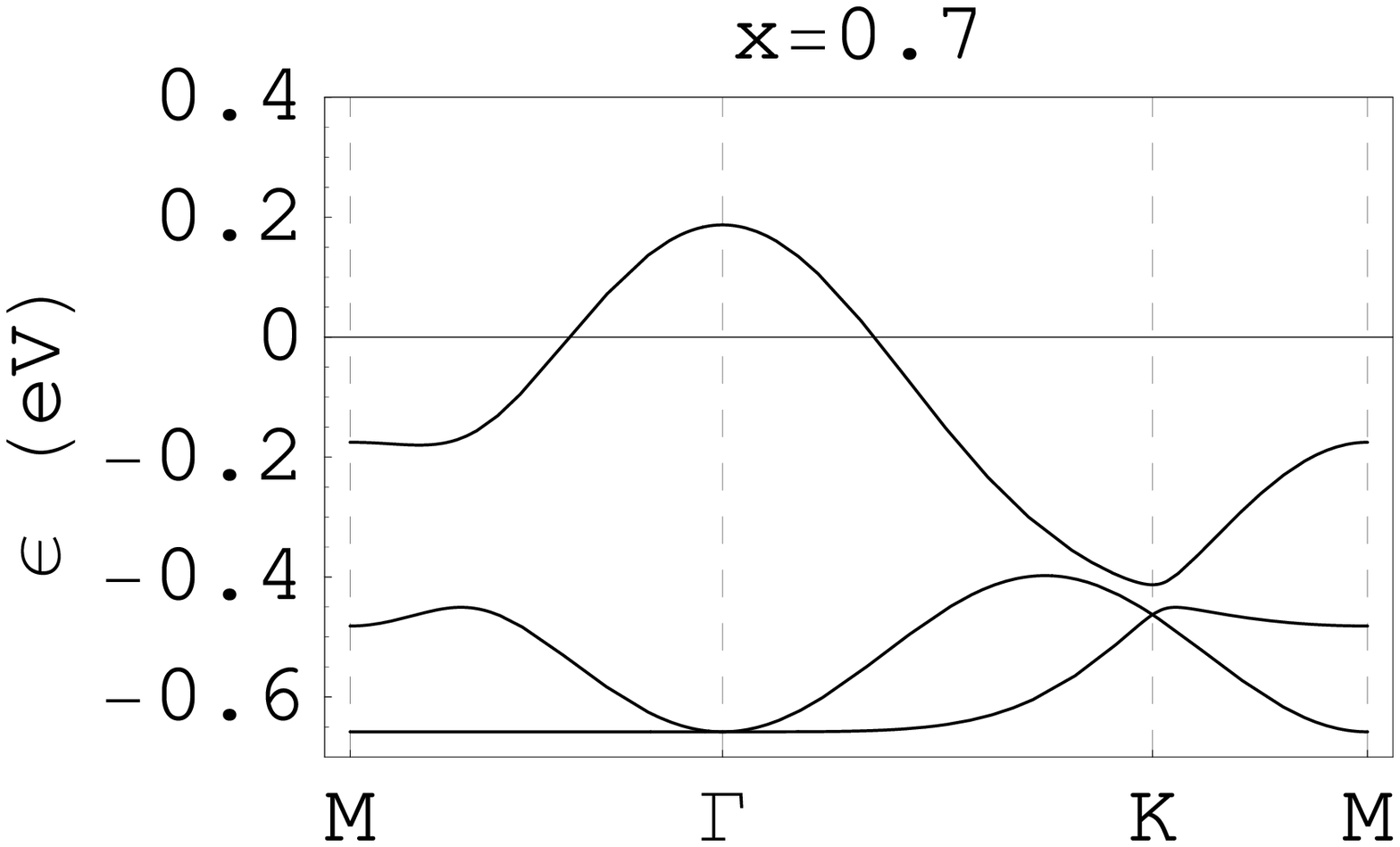}
	\includegraphics[width=.5\linewidth]{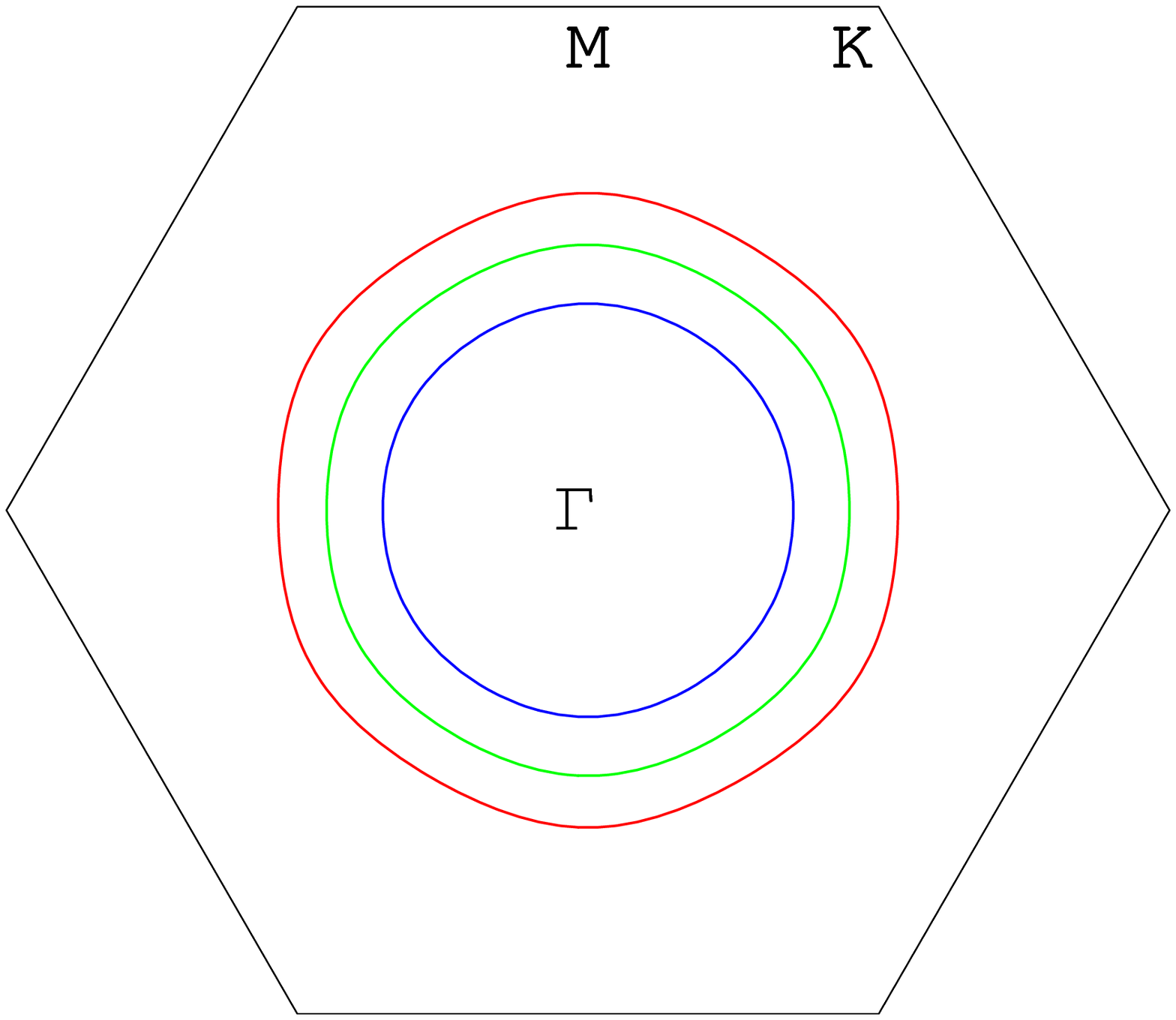}
	\caption{(Color online) Renormalized band-structures for dopings $x = 0.3,\ 0.5,\ 0.7$ and corresponding FS in red, green, blue respectively (larger FS corresponds to lower doping $x$).}
	\label{Fig:Results}
\end{figure}

In Fig.~\ref{Fig:Results} we show the resulting dispersion relations for electrons for three different values of $x$ near to those measured by ARPES, as well as the corresponding FS. As observed by ARPES~\cite{PRL95146401}, the feature that would give rise to hole-pockets near the $K$ points remains under $\epsilon_{F}$ with very little doping dependence, and the FS consists only in a central lobe around $\varGamma$ which verifies the Luttinger theorem as experimentally observed.

In summary, we have derived an effective multiband Hamiltonian $H_{eff}$ whose parameters are determined by cluster calculations. Without any new adjustable parameter, this model is able to reproduce the main features of the ARPES experiments.


\begin{thebibliography}{00}

\bibitem{NAT42253}
	K.~Takada et al.,
	\emph{Nature} \textbf{422} 53 (2003)
	
\bibitem{NAT423425}
	Y.~Wang et al.,
	\emph{Nature} \textbf{423} 425 (2003)

\bibitem{PRL96107201}
	J.~Bobroff et al.,
	\emph{Phys. Rev. Lett.} \textbf{96} 107201 (2006)

\bibitem{PRB70045104}
	K.W.~Lee et al.,
	\emph{Phys. Rev. B} \textbf{70} 045104 (2004)

\bibitem{PRL95146401}
	H.B.~Yang et al.,
	\emph{Phys. Rev. Lett.} \textbf{95} 146401 (2005)
	
\bibitem{PRL94146402}
	W.B.~Wu et al.,
	\emph{Phys. Rev. Lett.} \textbf{94} 146402 (2005)

\bibitem{Kroll1}
	T.~Kroll et al.,
	\emph{Phys. Rev. B} \textbf{74} 115123 (2006)

\bibitem{Kroll2}
	T.~Kroll et al.,
	\emph{Phys. Rev. B} \textbf{74} 115124 (2006)
	
\bibitem{PRL91257003}
	W.~Koshibae et al.,
	\emph{Phys. Rev. Lett.} \textbf{91} 257003 (2003)
	
\bibitem{CM0605454}
	S.~Landron et al.,
	condmat/0605454

\bibitem{PRL571362}
	G.~Kotliar et al.,
	\emph{Phys. Rev. Lett.} \textbf{57} 1362 (1986)

\bibitem{PRB475095}
	V.~Dorin et al.,
	\emph{Phys. Rev. B} \textbf{47} 5095 (1993)

\end{thebibliography}
\end{document}